# Aligned copper nanorod arrays for highly efficient generation of intense ultra-broadband THz pulses

S. Mondal[1], Q. Wei[1], W. J. Ding[2], H. A. Hafez[1], M.A. Fareed[1], A. Laramée[1], X. Ropagnol[1], G. Zhang[1], S. Sun[1], Z. M. Sheng[3,4,5], J. Zhang[4,5], and T. Ozaki[1*]

[1]Institut national de la recherche scientifique – Centre Energie, Matériaux et Télécommunications (INRS-EMT), 1650 Lionel-Boulet, Varennes, Québec J3X 1S2, Canada

[2]A*STAR Institute of High Performance Computing, Singapore 138632

[3]SUPA, Department of Physics, University of Strathclyde, Glasgow G4 0NG, UK

[4]Laboratory for Laser Plasmas and Department of Physics and Astronomy, Shanghai Jiao Tong University, Shanghai 200240, China

[5]Collaborative Innovation Center of IFSA, Shanghai Jiao Tong University, Shanghai 200240, China

[*]E-mail address: ozaki@emt.inrs.ca

We demonstrate an intense broadband terahertz (THz) source based on the interaction of relativistic-intensity femtosecond lasers with aligned copper nanorod array targets. For copper nanorod targets with length 5 µm, a maximum 13.8 times enhancement in the THz pulse energy (in ≤ 20 THz spectral range) is measured as compared to that with a thick plane copper target under the same laser conditions. A further increase in the nanorod length leads to a decrease in the THz pulse energy at medium frequencies (≤ 20THz) and increase of the electromagnetic pulse energy in the high-frequency range (from 20 - 200 THz). For the latter, we measure a maximum energy enhancement of 28 times for the nanorod targets of length 60 µm. Particle-in-cell simulations reveal that THz pulses are mostly generated by coherent transition radiation of laser produced hot electrons, which are efficiently enhanced with the use of nanorod targets. Good agreement is found between the simulation and experimental results.

**PACS numbers:** 42.79.Nv, 52.59.Ye, 52.25.Os, 52.38.Kd

Terahertz radiation is a powerful tool for imaging [1] and probing various physical systems [2], because of its unique nonionizing nature and transparency to many materials [3].

It also has strong interactions with diverse materials, which allow THz radiation to probe many physical, chemical and biological systems [4–7]. Intense broadband THz pulses are opening new scientific areas to explore, such as nonlinear optics in the THz domain [8], as well as new technological opportunities, such as single-shot THz spectroscopy and imaging. Stimulated by such applications, the generation of intense THz pulses has recently been the subject of great interest, and several tabletop techniques have been studied in detail [9,10]. For example, optical rectification sources and THz generation from air plasma [11,12] have been the subject of many extensive efforts to scale up both THz peak electric field and energy [9]. THz pulses with $E_{peak\_electric\_field}$ > 100 MV/cm have been demonstrated by difference-frequency mixing [13]. Electron accelerators can also be used to produce energetic THz pulses but usually they are not available at modest laboratory scales [14,15]. Progress in research up to now has made it clear that many laser-based methods possess an upper limit in the maximum driving laser intensity [16] that could be used, which is limiting the THz yield and hence their applications to several spectroscopic purpose [4,17].

High-intensity ultrafast laser-plasma interaction can also be a potential source of intense ultra-broadband THz radiation [18–21]. Particle-in-cell (PIC) simulations have predicted that GV/cm THz fields can be generated by such interactions via different mechanisms such as laser wakefield excitation in underdense plasma [22] and coherent transition radiation (CTR) with solid targets [23]. Intense transition radiation (TR) in the THz frequency range has been observed when energetic electrons generated by high-intensity laser-solid interaction leave the target [23,24]. THz pulses (0.1 – 133 THz) with pulse energy of > 700 µJ have been demonstrated from rear side of a foil target via a different mechanism, THz generation via target normal sheath acceleration [21,25]. Intense THz pulses via CTR generation in the rear side of size limited thin targets have also been demonstrated by Liao *et al.* [24]. Generated via the energetic electrons, transition radiation could reveal the characteristics of those electron bunch produced during the laser-solid interaction and hence is very important in high-intensity laser plasma interaction [26,27].

In this work, we first study TR in the THz frequency range generated by relativistic intensity laser-plasma interaction using thick copper (Cu) targets. Then, we demonstrate that aligned Cu nanorod targets, which have been used to improve the efficiency of generating highly energetic photons and particles [28–34], can also be used to significantly enhance the THz pulse energy through transition radiation. Single-shot electro-optic measurement shows the temporal behavior as well as the coherent nature of the THz pulses generated via TR.

Figure 1*(a)* shows a schematic diagram of the experimental setup. We use 10 TW femtosecond laser pulses with high contrast (nanosecond contrast ~$10^{-7}$) from the 10 Hz beam line of the Advanced Laser Light Source (ALLS) facility at INRS-EMT. This laser, with incidence angle of 45° and p-polarization, is focused on to a Cu target (size: 5 cm × 5 cm × 3 mm) using an f/3 off-axis parabolic mirror, to a circular spot of 20 µm in diameter. The target is mounted on an automated XYZ translation stage inside a vacuum chamber. The 10 Hz beam line can deliver up to 240 mJ maximum pulse energy with 40 fs pulse duration after compression, with a central wavelength of 800 nm, which when focused results in a peak intensity of about $3.5 \times 10^{18}$ Wcm$^{-2}$. Ultra-broadband THz and infrared (IR) pulses (via CTR generation) are generated as a result of the laser-plasma interaction. THz pulses are then collimated by using a thick gold plated off-axis parabolic mirror, and then guided out of the vacuum chamber through a THz window made from ultrahigh molecular weight polyethylene (UHMWPE), which only transmits radiation up to 20 THz while blocking higher frequency electromagnetic radiation. THz pulses are then refocused on to a calibrated pyroelectric

detector (Gentec-EO, THZ5I-BL-BNC) using another off-axis parabolic mirror to measure THz pulse energy. To remove any residual of the 800 nm driving laser, we have further used two high-resistivity float-zone silicon (HRFZ-Si) filters that have flat transmission up to the cut-off at around 1.5 µm. We take an average over 15 shots to improve the statistical error. The THz pulse energy from the optically polished target is compared with that from the aligned Cu nanorod targets. THz pulse energy is collected over a relatively small solid angle of 0.0873 Sr.

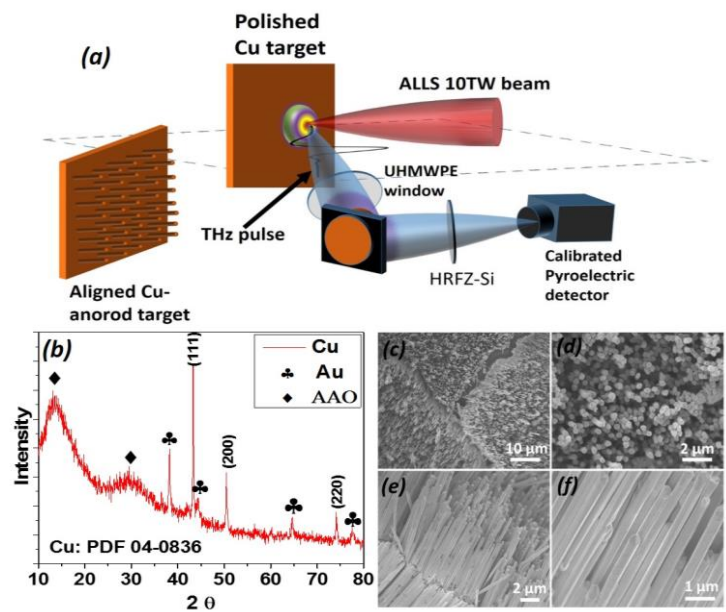

*Figure 1: (Color online) (a) Schematic diagram of the experimental setup. (b) X-ray diffraction (XRD) spectrum of Cu nanorods embedded in AAO template. (c), (d), (e) & (f) SEM images of Cu nanorod arrays at different magnifications: (c, d) top view; (e, f) cross-sectional view.*

The commercially available porous anodic aluminum oxide (AAO) membrane (pore size: 200 nm, membrane thickness: 60 µm, Whatman® Anodisc) was used as a template for the electrochemical deposition of Cu nanorod arrays [35]. A gold layer (300 nm) was sputtered on one side of the through-hole AAO template, serving as the working electrode in a conventional three-electrode cell for Cu electrochemical deposition, with graphite carbon and saturated calomel electrode (SCE) electrode as the counter and reference electrode, respectively. The electrolyte was 0.2 M $CuSO_4 \cdot 5H_2O$ + 0.1 M $H_3BO_3$ for Cu deposition.

Experiments were carried out using Potentiostat (Autolab) with the constant potential of -1.20 V (vs. SCE) at room temperature. The length of the Cu nanorods can be controlled between 0 and 60 μm by adjusting the deposition time.

The crystallographic studies of Cu nanorods embedded in AAO were carried out by X-ray spectroscopy (XRD, Bruker D8 Advanced Diffractometer, Cu Kα radiation). The XRD spectrum of the Cu nanorods, shown in Fig. 1*(b)*, fits the standard XRD pattern very well. Three reflection peaks attributed to (111), (200), (220) are evidently noticeable, and can be completely indexed to the Cu face-centered cubic crystal structure (JCPDS 04-0836) [36]. There are no impurities except for the broad peaks belonging to the amorphous AAO template and the corresponding peaks of the sputtered Au electrode.

For the SEM characterization, the as-prepared Cu nanorods embedded in AAO template were first immersed in the NaOH solution to dissolve the alumina membrane, and then the nanorods were washed thoroughly with distilled water and ethyl alcohol several times. Figs. 1*(c)*, 1*(d)*, 1*(e)* and 1*(f)* show the SEM images of the Cu nanorods (length $h$ = 15 μm) at different magnifications. It is clearly depicted from the SEM micrograph that a large quantity of well-aligned, dense homogeneous in diameter, and are parallel to each other Cu nanorods have been successfully fabricated by this technique.

In Fig. 2*(a)*, we compare the THz pulse energy for medium frequencies (≤ 20 THz, with UHMWPE window) detected by the pyroelectric detector for polished and rough Cu thick-solid targets. The presence of small random structures on the target surface could increase the THz pulse energy by about 4 times. In Fig. 2*(b)*, we show the THz pulse energy emitted when high-intensity femtosecond laser irradiates the aligned Cu nanorod targets with a nanorod diameter (*D*) of 200 nm and lengths (*h*) of 5 μm, 15 μm, 30 μm, 40 μm), and compare them with the THz pulse energy emitted from an optically polished Cu target.

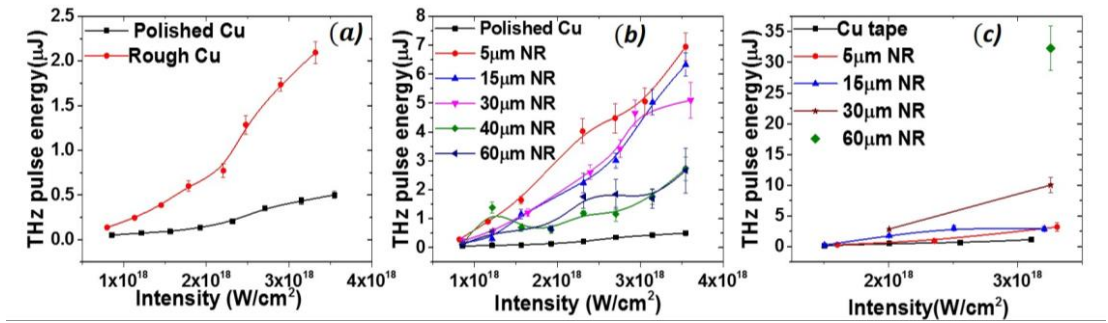

*Figure 2: (Color online) THz pulse energy for different targets as a function of intensity on target. (a) Comparison between thick polished and rough Cu solid targets in spectral range ≤ 20THz; (b) Comparison between nanorod (NR) targets of different nanorod lengths (h) in spectral range ≤ 20THz; (c) THz pulse energy for different nanorod (NR) lengths, with high-pass window in spectral range from 20THz to 200THz.*

Compared with polished Cu targets, we observe a 13.8 times enhancement in the THz pulse energy for *h* = 5 μm nanorod targets. However, we also find that the THz pulse energy decreases as nanorod length *h* increases further. This is in contrast to earlier studies on X-ray generation, which shows enhancement in the X-ray yield with nanorod length [31]. To investigate this difference, we repeat the experiment with a high-pass window that has lower transmission for lower frequencies (1% for ≤ 20 THz, ~10% transmission for ≥ 100 THz and ~60% for ≥ 130 THz). Radiation with frequencies higher than 200 THz is blocked by the HRFZ-Si filters. The results are shown in Fig. 2*(c)*. In this measurement, the reference material is Cu tape, because the Cu nanorod targets were attached to the bulk Cu target by Cu tape, which has adhesives on both sides. For higher frequencies of electromagnetic radiation (20 THz to 200 THz), the pulse energy increases with the nanorod length (*h*), and the enhancement in the pulse energy reaches as high as 28 times for the case of the 60-μm-long Cu nanorod target, demonstrating a large pulse energy of 32 μJ per pulse in 0.0873 Sr solid angle.

The THz pulses, especially for low frequency THz radiation, are emitted in a broad angle, which is also seen in the simulation results shown later. However, we only collect THz signal over a small solid angle of 0.0873 Sr, and thus we obviously do not collect all of the THz energy generated by the interaction. To estimate the total THz pulse energy generated by the interaction, we consider that the THz pulses are emitted in a $2\pi$ solid angle [21,37] on the target front. By correcting for the small solid angle of detection, we estimate that several 100 µJ of THz energy is generated at frequency range (≤20 THz). Most of this THz pulse energy can easily be collected by using an ellipsoidal mirror [21], which opens a roadway towards millijoule (mJ) class THz sources on tabletop.

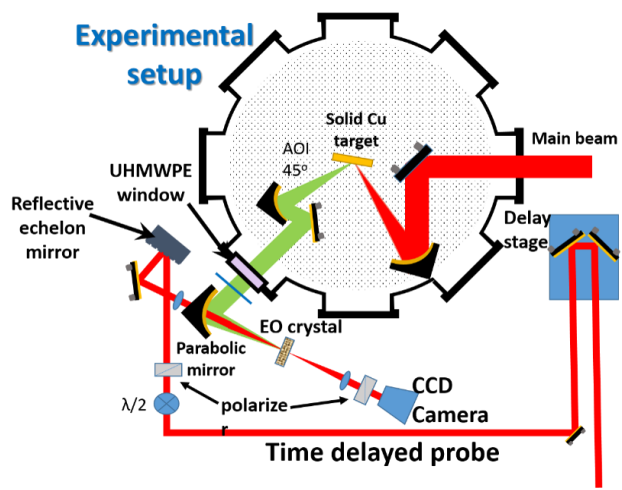

*Figure 3: Schematic setup for single-shot electro-optic measurement of THz pulses.*

Temporal profile of the THz pulses have been measured by using single-shot electro-optic (EO) measurement using a reflecting echelon mirror, a technique similar to the one used in Ref. [38]. The experimental scheme is shown in Fig. 3. Details of single-shot electro-optic measurement of THz pulses are described in the supplementary material.

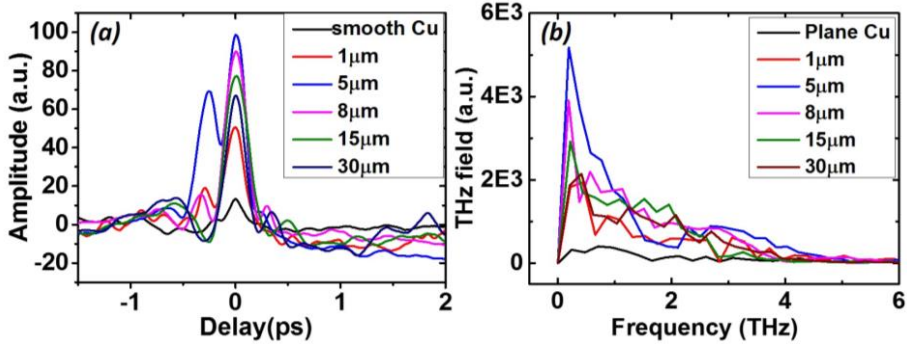

*Figure 4: Comparison of THz pulses (a) temporal domain (b) in frequency domain captured in single-shot experiment with polished and nanorod targets.*

The THz field profile in temporal domain as well as frequency domain obtained from polished and nanorod targets are shown in Fig. 4*(a) and* 4*(b).* The temporal profile of the THz pulses varies from shot to shot, and so in Fig. 4*(b),* we show the data that has the maximum THz peak field out of 20 successive shots. A simple Gaussian fit provide a pulse width (FWHM) of 235±6 fs for the THz pulse obtained from 8μm nanorod target, which is expected by the ultra-broadband nature of the THz pulses.

To understand our experimental results in detail, we have performed two dimensional particle-in-cell (PIC) simulations. The geometry of the simulation box is shown in Fig. 5(a). The simulation parameters are set close to those of the experiment. In the simulation, a p-polarized Gaussian beam with wavelength of 800 nm, pulse width of 40fs and intensity of 3.5× $10^{18}$ W/cm$^2$ is incident onto the target at an incidence angle of 30º. Since the prepulse contrast of the laser in the experiment is high, we have not included preplasma in the simulation. An 8 μm thick copper target is used, which is covered with Cu nanorod arrays with 200 nm diameter (*D*) and spacing (*$d_0$*) 200 nm. Figure 5(b) is a snapshot of the spatial distribution of the magnetic field, which is time-averaged over a laser period in order to filter out the high frequency components. Both backward and forward radiations are emitted from the target front and back sides, respectively, which can be attributed to coherent transition radiation generated by hot electrons produced in the laser-plasma interaction [23,39]. For polished planar targets, the THz radiation is the weakest around the specular reflection direction and strongest along the target surface direction, typical for transition radiations. The simulation shows that the nanorods have enhanced the backward THz radiation and changed its emission direction to the specular reflection direction, significantly from that predicted for a planar target. These are more obvious from the spectra of the electromagnetic fields, which is detected 100 μm away from the laser irradiated spot in the target front, as shown in Figs. 5(c) and 5(d).

To check the effect of nanorod length on THz radiation, we have varied the nanorod length from 0 to 30 μm, which shows the presence of optimized nanorod length for THz radiation. When the nanorod length increases from 0 to 4 μm, the radiation intensity decreases along the target surface (near 0° or 180°) while it increases in the specular reflection direction (120°). As a whole, it become dominant on the specular reflection direction. Figure 5(e) shows that the intensity of the backward THz radiation in the specular reflection direction varies with the nanorod length, where the intensities are normalized to the maximum THz emission from polished targets. As the length increases, the radiation intensity increases until $h$ = 4 μm, then it decreases slightly for $h$ > 4 μm, which reaches a saturation for $h$ > 10 μm. Because the emitted THz radiation is coherent transition radiation by hot electrons, the radiation is the strongest when the absorption rate of the laser energy by electrons is the highest. The corresponding optimal nanorod length, at which the highest number of hot electrons is produced, is determined by the nanorod parameters $d_0$ and $d_1$ [33]. Simulations show enormous increase in both the number and total energy of hot electrons for the case of nanorod targets, when compared with polished targets. The variation in the total kinetic energy of hot electrons as a function of the nanorod length (*h*) is shown in Fig. 5(*f*). The nanorod array enhances the electron energy in both the x ($E_{kx}$) and y ($E_{ky}$) directions, but much more significantly in the y (vertical) direction. $E_{ky}$ peaks at around *h* = 3 μm, although the energy in the horizontal x-direction $E_{kx}$ is not the maximum. The electron energy in the y-direction $E_{ky}$ contributes more to THz emission in specular reflection direction (120º) than the electron energy in the x-direction ($E_{kx}$). This is consistent with the theory of coherent transition radiation that the radiation is mostly emitted in a large angle or nearly perpendicular to the moving direction of the electron [23,39,40] when the electron is in moderate energy. The average kinetic energy of hot electrons is normally at the order of 100 keV under the laser conditions. The maximum $E_{ky}$ is about 3.4 times higher than that from polished targets. The tendency of the THz intensities agrees very well with the behavior of the total energy of hot electrons in the vertical direction.

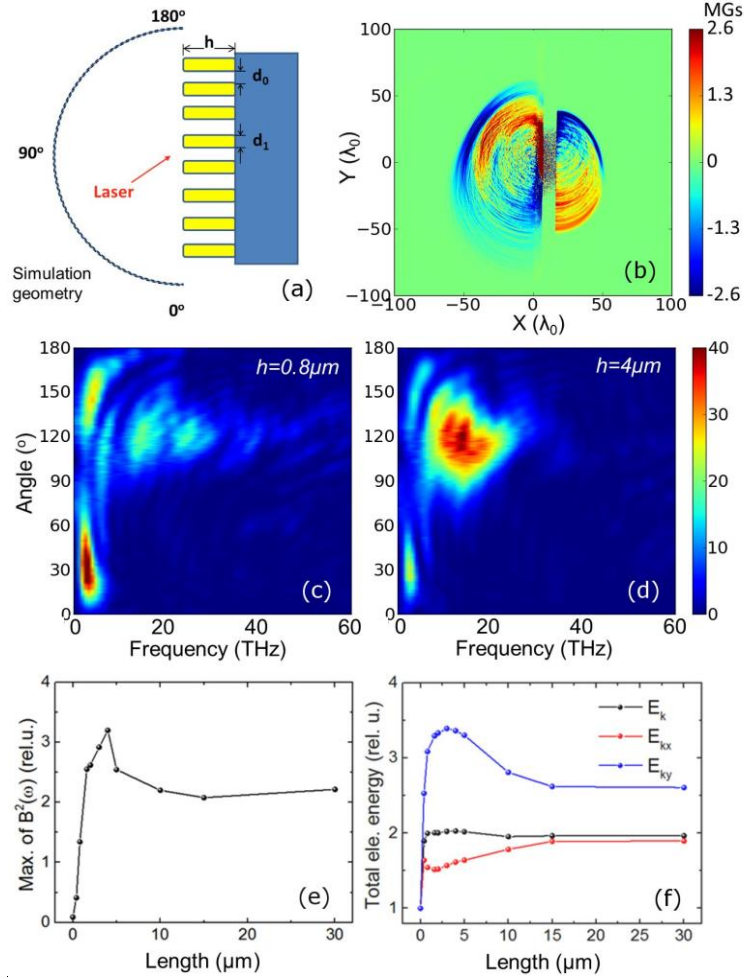

*Figure 5: PIC simulation results: (a) Simulation geometry. (b) Snapshot of magnetic field $B_z$ (averaged in a laser cycle) of the THz radiation. (c) and (d) Spectra of EM fields emitted from the target front with the nanorod length is 0.8 µm and 4 µm, respectively. (e) Intensity of THz radiation (maximum in THz range of the spectra) as a function of the nanorod length. Radiations are detected in the reflection direction (120º). (f) Total kinetic energy of hot electrons ($E_k$ > 30 keV) as a function of the nanorod length. Energies are rescaled to that of planar target.*

In summary, we have demonstrated that the THz pulse energy can be increased by using aligned Cu nanorod array targets, with a maximum enhancement of up to 13.8 times for ≤ 20 THz and 28 times in the spectral range between 20 THz and 200 THz. It is shown that there is an optimal nanorod length for most efficient THz radiation. The temporal profiles of the THz pulses are also obtained by use of single-shot EO measurement. PIC simulations reveal the mechanism behind the intense THz pulse generation by this technique, which we attribute to CTR at THz frequencies by the energetic electrons produced by the laser plasma interactions. It also reveals that the THz radiation with nanorod array targets is distributed mainly in the specular reflection direction, different from that with a planar target.

T. O. and S. S. acknowledge the financial support from the Fonds de recherche du Québec – Nature et technologies (FRQNT) and the Natural Sciences and Engineering Research Council of Canada (NSERC). Z.M.S. and J.Z. acknowledges the support by the National Basic Research Program of China (Nos. 2014CB339801 and 2013CBA01504), the National Science Foundation of China (No. 11421064), a MOST international collaboration project (No. 2014DFG02330), and a Leverhulme Trust Research Grant.